\documentclass[prb,aps,twocolumn,notitlepage]{revtex4-1}

\usepackage{graphicx}
\usepackage{ifthen}
\usepackage{ifpdf}
\usepackage{color}
\usepackage{latexsym}
\usepackage{amsmath}
\usepackage{amssymb}
\usepackage{bm}
\newcommand{\half}{\mbox{\small $\frac{1}{2}$}}

\newcommand{\eexp}{\mbox{e}^}

\usepackage{bbm}

\newcommand{\beq}[1]{\begin{eqnarray}\ifthenelse{#1=-1}{\nonumber}
{\ifthenelse{#1=0}{}{\label{e#1}}}}
\newcommand{\eeq}{\end{eqnarray}}
\newcommand{\hide}[1]{}

\begin{document}
\title{Electron spin spectroscopy of single TEMPO dimers coupled via interfering tunneling currents}
\author{Yishay Manassen$^{1,2}$, Moamen Jbara$^{1,3}$, Michael Averbukh$^1$, Zion Hazan$^1$, Carsten Henkel$^4$ and Baruch Horovitz$^{1,2}$}
\affiliation{$^1$Department of Physics, Ben Gurion University of the Negev, Beer Sheva 84105, Israel\\
$^2$ The Ilse Katz Institute for Nanoscale Science and Technology, Ben-Gurion University of the Negev, Beer Sheva 84105, Israel \\
$^3$ Department of Chemistry, Technion, Haifa, Israel 32000 \\
$^4$ University of Potsdam, Institute of Physics and Astronomy, 14476 Potsdam, Germany}
\begin{abstract}
We report the detection of electron spin resonance (ESR) in individual dimers of the stable free radical 2,2,6,6-tetramethyl-piperidine-1-oxyl (TEMPO). ESR is measured by the current fluctuations in a scanning tunnelling microscope (ESR-STM method). The multi-peak power spectra, distinct from macroscopic data, are assigned to dimers having exchange and Dzyaloshinskii-Moriya interactions in presence of spin-orbit coupling. These interactions are generated in our model by interfering electronic tunneling pathways from tip to sample via the dimer's two molecules.
This is the first demonstration that tunneling via two spins is a valid mechanism of the ESR-STM method.
\end{abstract}
\maketitle

The attempt to detect and manipulate a single spin in individual molecules is a fundamental challenge \cite{wrachtrup,koehler,rugar,elzerman,xiao}. A promising tool to monitor the electron spin resonance (ESR) on the nm-scale is based on a scanning tunneling microscope (STM) that measures current-current correlations in a static magnetic field (ESR-STM) \cite{manassen1,manassen2,manassen3,durkan1,durkan2,komeda,saaino}, rather than using external radiofrequency fields. The experiments so far resulted in a signal at the Larmor frequency, a signal that is sharp even at room temperature and whose frequency varies linearly with the applied magnetic field \cite{durkan1,komeda}. Such a current spectrum has been observed in several spin systems, including dangling bonds \cite{manassen1,manassen2}, metal impurities in silicon \cite{manassen3} and adsorbed paramagnetic molecular radicals. \cite{durkan1,durkan2} On the Si(111)$7\times7$ surface, two peaks show up and relate to defects that differ in STM images \cite{komeda,saaino}. ESR-STM has been used to detect the hyperfine spectrum of a single spin in SiC \cite{balatsky2,manassen5}. Similarity to macroscopic ESR was demonstrated in the spectrum of silicon vacancy \cite{manassen5}, showing hyperfine contributions from $^{29}$Si nuclei.

Recently, a different type of ESR-STM was observed at low temperatures, using a spin polarized tip and rf irradiation \cite{mulleger,baumann,willke,seifert}. Furthermore,                                                                                                                                                                                                  single spin ENDOR (electron nuclear double resonance) was performed \cite{manassen6} by applying an rf field at frequencies of the nuclear transitions and monitoring the intensity of the hyperfine line observed by ESR-STM; this facilitated measurements of the hyperfine coupling, the quadrupole coupling and the nuclear g-factors.

Several theoretical models for ESR-STM have been put forward \cite{caso,golub,balatsky1,manassen4,horovitz1},
and it was shown that tunneling via a single spin cannot explain the observations \cite{caso,golub}.
Instead there must be at least two tunneling channels whose interference generates ESR-STM \cite{caso,golub,horovitz1}. The two channels are most likely due to two distinct spin sites: one is the target scanned by the STM probe, while the other is possibly located on the tip itself. The lack of direct evidence for the second spin has made elusive the interpretation of the experiments.

In the present work we consider a system of 2,2,6,6-tetramethyl-piperidine-1-oxyl (TEMPO) molecules that agglomerate and are likely to form dimers, see Fig.\;1. The TEMPO molecule is a stable free radical carrying a spin 1/2, hence a dimer would be an ideal setup for forming parallel tunneling routes that lead to the ESR-STM phenomenon.
In our experiment, neither tip nor substrate are spin polarized and rf radiation is not applied.
We model the data with a two-spin model that allows for exchange and Dzyaloshinskii-Moriya interactions \cite{horovitz1,horovitz2}. We find reasonable agreement of the theory with many experimental spectra, thus providing a complete theoretical interpretation. We note also that the dimer scenario is of much interest to quantum information science, since by tuning parameters, a long-lived dark state is available and quantum entanglement can be achieved \cite{horovitz2}.

In the ESR-STM experiments we performed, the molecules are deposited on gold films of thickness 100nm on Mica \cite{buchholz}. TEMPO is dissolved in toluene and drop casted on the surface at a concentration of $0.041\,{\rm g}/25\,{\rm ml}$, corresponding to one monolayer. After drop casting the sample is put in a UHV chamber (pressure range $10^{-10}\,{\rm torr}$), and tunneling probe data are taken at room temperature. The clean Au surface shows flat terraces of variable shape, some of them triangular (Fig.\;1a).
After deposition, the molecules disperse as individual entities fairly uniformly on the surface (Fig.\;1b), afterwards there is a slow agglomeration process (several days) where some parts of the surface end up clean gold (as in Fig.\;1a), while in other parts there is a dense monolayer of TEMPO molecules (Fig.\;1c). STM studies of TEMPO adsorbed on Si(111)$7\times7$ have shown that the molecule adsorbs with its NO axis (Fig.\;1d) normal to the surface \cite{pitters}. More relevant here are studies of TEMPO with Au spheres, showing the disappearance of the ESR signal when the NO group is close to the Au surface \cite{zhang}.

\begin{figure}[htb]
\includegraphics*[width=.4 \columnwidth]{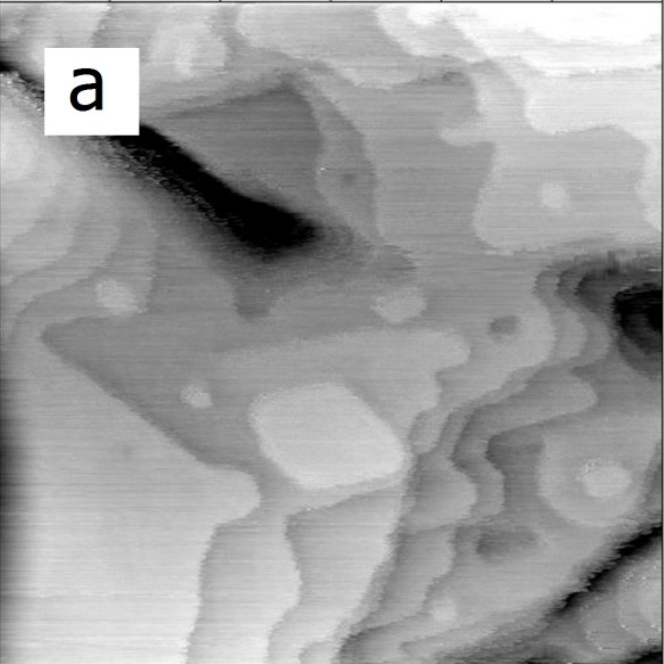}
\includegraphics*[width=.4 \columnwidth]{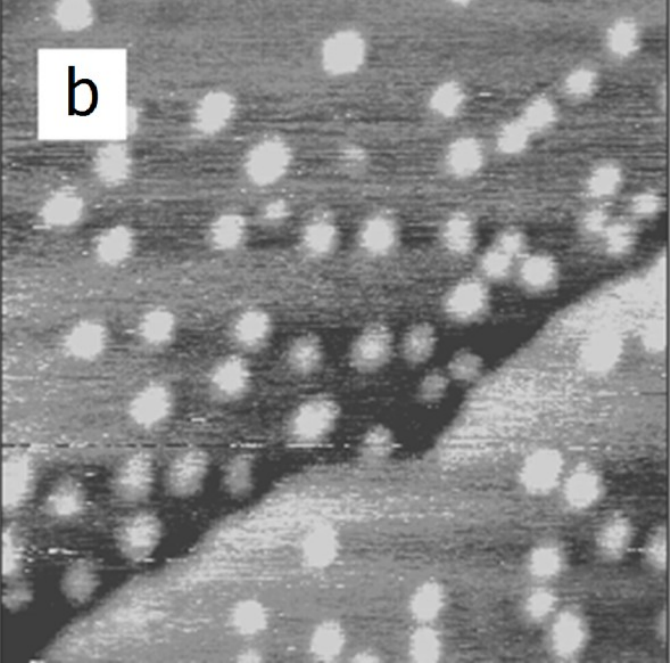}
\includegraphics*[width=.4 \columnwidth]{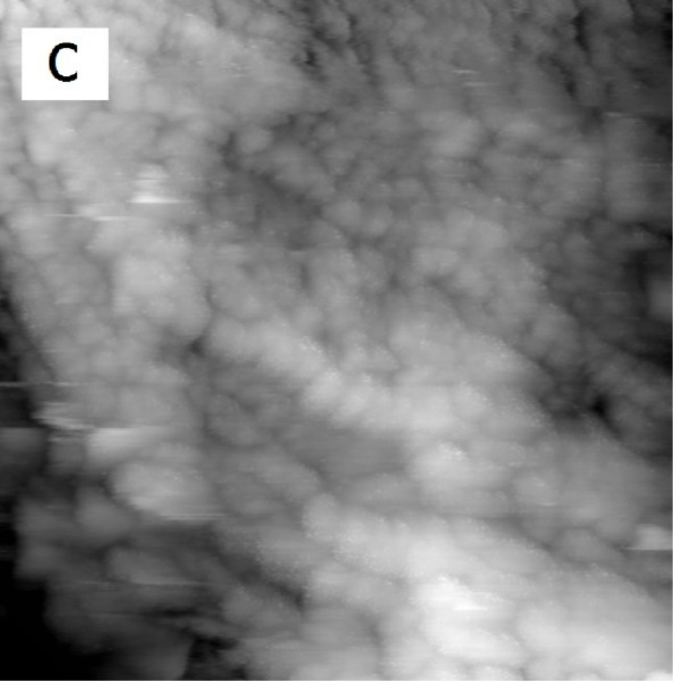}
\includegraphics*[width=.4 \columnwidth]{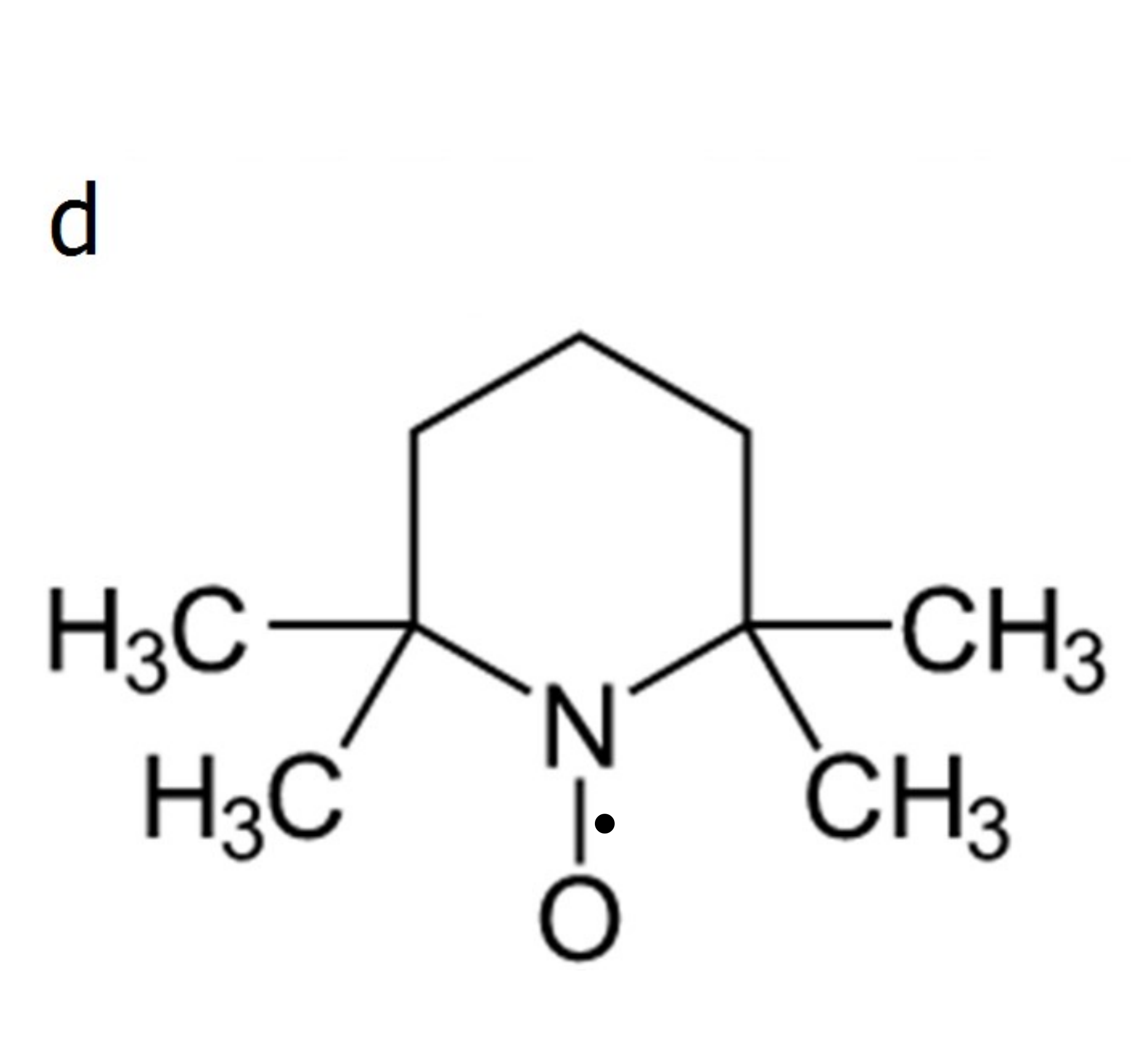}
\caption{STM image of Au(111) surface before deposition of TEMPO. (b) Au(111) after deposition with single molecules. (c) TEMPO adsorbed on Au(111) when agglomeration is reached. All images are $30\times30\,{\rm nm}^2$.
Tunneling conditions: (a) $-1\,{\rm V},\, 0.4\,{\rm nA}$; (b) $-0.1\,{\rm V},\, 0.5\,{\rm nA}$; (c) $-0.5\,{\rm V},\, 0.5\,{\rm nA}$. (d) The molecular structure of TEMPO. The unpaired electron occupies a $p_z$ orbital on the N$-$O 
bond \cite{kobayashi}.}
\end{figure}

The g-tensor and hyperfine coupling of TEMPO are well documented \cite{kobayashi}, showing an almost isotropic g-factor $g \approx 2.007$ (within .5\%). The eigenvalues of the hyperfine interaction with the nuclear ${}^{14}{\rm N}$ spin are 17, 15, and 94 MHz for the $x$, $y$ and $z$ directions, respectively (the singly-occupied p-orbital of N defines the $z$-axis). The dominant hyperfine coupling $a$ in the direction parallel to the magnetic field depends on the molecular orientation and possibly on averaging due to rotational motion. Hence $a$ is a fitted parameter that turns out to be $50{-}80\,{\rm MHz}$ between various molecular sites. The nominal external magnetic field, perpendicular to the surface, is $230\,{\rm G}$, which with $g=2$ would correspond to a Larmor frequency of $\nu=644\,{\rm MHz}$, yet we take $\nu$ as a fitting parameter, allowing for uncertainties in the actual field.
Experimental spectra of the tunnel current are taken in 9800 channels covering the range of 580-780\,MHz, each channel requiring about $50\,\mu{\rm s}$ acquisition time. This is short compared to the spin lifetime of the $^{14}$N nucleus, typically in the range of $0.5{-}1\,{\rm ms}$ [Ref.\,\onlinecite{odell}], so that we may assume fixed spin states. The spectrum analyzer averages 200 spectra, however, so that eventually nuclear spin flips occur, and an ensemble of all levels is probed. A whole spectrum of a single site takes 90\,s.
%

We first outline our model and results. Each TEMPO molecule has three $^{14}{\rm N}$ nuclear spin states whose hyperfine coupling to the unpaired electron spin splits the ESR into $\nu,\,\nu \pm a$. A dimer with two spins has then nine hyperfine states with nuclear spin projections $m,\,m'= 0,\pm 1$. When electrons transit between the electrodes (tip and substrate) through two TEMPO molecules, interfering exchange tunneling events occur and generate interactions between the two molecular spins as well as dissipation (linewidth $\Gamma$) \cite{horovitz1,horovitz2}. Both interactions and dissipation depend sensitively on the energy levels of the two molecules being of two types, either degenerate ($m=m'$) or non-degenerate ($m\neq m'$). We assume that $a \gg \Gamma$ (confirmed by our analysis) so that the distinction between the two types is well defined. There are three dimer states with $m=m'$ that are degenerate and therefore have more (secular) terms to be kept when deriving the master equation; these generate exchange as well as Dzyaloshinskii-Moriya interactions \cite{horovitz2}.
In addition there are six non-degenerate pairs $m\neq m'$ with a weaker exchange interaction that splits the three ESR transitions into six peaks. Although the degenerate pairs give more peaks, their statistical weight is smaller,
and only the two extreme ones are usually visible, leading to overall eight peaks. It is interesting to note that the behaviour in the more prominent dips between the peaks is most informative: their positions determine the bare hyperfine parameter, and their shapes are sensitive to couplings among degenerate pairs.

We proceed to describe our theoretical model for TEMPO dimers in more detail.
Compared to its component $a$ parallel to the magnetic field, the transverse hyperfine splitting $b$ can be neglected because it affects the spectra only in second order $\sim (b/\nu)^2$. Such terms may cause a small difference between the two hyperfine splittings and even shift $\nu$.
In addition to Larmor and hyperfine terms, the two-spin Hamiltonian contains effective interactions due to tunneling to either tip or substrate. They were derived in Refs.~\onlinecite{horovitz1,horovitz2} within a Born-Markov master equation and turn out fairly large, as they depend on the electron bandwidth $\Delta$ of the electrodes. The degenerate cases (nuclear levels $m = m'$) involve exchange coupling ($J_{\rm ex}$) and Dzyaloshinskii-Moriya coupling ($J_{\rm DM}$). The latter appears only when spin-orbit coupling in the tunneling junction is taken into account which
is actually essential to the presence of ESR-STM \cite{horovitz1}. It is parametrized by two angles $\theta, \phi$ of an SU(2) spin rotation matrix acting on the tunneling electron spin.
The interaction strengths are
\beq{01}
\left.\begin{array}{r}
J_{\rm ex}\\
J_{\rm DM}
\end{array}
\right\} &=& 4J_1J_2N^2(0)\Delta\cos\half\theta
\left\{ \begin{array}{r} \cos\phi\\ \sin\phi \end{array} \right.
\eeq
where $J_1,\,J_2$ are the exchange tunneling elements via the two localised spins, respectively, and $N(0)$ is the density of states of either electrode at its Fermi level. The Hamiltonian of a degenerate dimer, for $m=-1,\,0,\,1$, is
\beq{02}
{\cal H}_{\rm deg}&=&\half(\nu+am)[\tau_z\otimes\mathbbm{1}+\mathbbm{1}\otimes\tau_z]
-J_{\rm ex}\bm\tau\otimes\bm\tau
\nonumber\\
&& {} + J_{\rm DM}[\tau_x\otimes\tau_y-\tau_y\otimes\tau_x]
\eeq
where the tensor products describe the two localized spins. The exchange tunneling between the tip and substrate, including the spin-orbit coupling, also leads to a finite linewidth for each spin, $\Gamma_1,\,\Gamma_2$.
We note that in this model the dimer molecules need not to be close to each other, since the interactions are generated via tunneling to tip or substrate. Indeed, the separation of molecules in Fig.\;1b is 2-4\,nm, a scale on which the direct dipole-dipole interaction can be neglected, justifying our model Hamiltonian.

The non-degenerate dimer states are decoupled from the degenerate ones in the master equation by the secular approximation, valid in our experiment where $a \gg \Gamma_{1,2}$. Each pair $m\neq m'$ has an anisotropic exchange coupling (being the only secular term \cite{horovitz2}), and the Hamiltonian is
\beq{03}
{\cal H}_{\rm nondeg}&=&\half(\nu+am)\tau_z\otimes\mathbbm{1}+\half(\nu+am')\mathbbm{1}\otimes\tau_z
\nonumber\\
&&
{} -J_{\rm ex}\tau_z\otimes\tau_z
\eeq
It is useful to list the eigenstates, levels and transition frequencies, see Table\,I. We note that the hyperfine transitions $\nu+am,\,m=-1,\,0,\,1$ are split by the interactions. In particular $J_{\rm DM}$ is responsible for the splitting of the $T1\rightarrow T2,\, T2\rightarrow T3$ transitions.

\begin{table*}[t]
\caption[]{Spectra and transitions of degenerate dimers (first 4 lines where $\tan\psi=J_{\rm DM}/J_{\rm ex}$, $J = \sqrt{J_{\rm ex}^2+J_{\rm DM}^2}$ and nuclear levels $m = m' = -1,\,0,\,1$) and non-degenerate dimers (last line where $s,s'=\pm 1$ are electron spin states). The $S,\,T$ notation refers to singlet or triplet states that decouple exactly in  the limit $J_{\rm DM}=0$.
Transitions are given only near the Larmor frequency $\nu$.
}
\begin{tabular}{|l|c|c|}
\hline
  eigenstates  &  energy levels &  transition frequencies
\\ \hline
$S:  |{\uparrow\downarrow}\, mm\rangle-\eexp{i\psi}|{\downarrow\uparrow}\, mm\rangle$ & $2J+J_{\rm ex}$ & $S\rightarrow T1,\,T3: $
\\
$T1: |{\uparrow\uparrow}\, mm\rangle$ & $\nu- J_{\rm ex}+am$ &  $(\nu+am)\pm 2(J+J_{\rm ex})$
\\
$T2:  |{\uparrow\downarrow}\, mm\rangle+\eexp{i\psi}|{\downarrow\uparrow}\, mm\rangle $ & $-2J+J_{\rm ex}$ & $T1\rightarrow T2,\, T2\rightarrow T3:$
\\
$T3: |{\downarrow\downarrow}\, mm\rangle$ & $-\nu-J_{\rm ex}-am$ & $\nu+am \pm 2(J-J_{\rm ex})$
\\
\hline
$|ss'm\neq m'\rangle$
& $\half s(\nu + am) + \half s'( \nu + a m') - ss' J_{\rm ex}$
& $\nu+am\pm 2J_{\rm ex}$
\\
\hline
\end{tabular}
  \label{tab:1}
\end{table*}

The fluctuations in the tunneling current arise from spin flips. We model these by time correlations of two spin operators, one carries a current with spin flip $\tau_{\pm}$, the other carrying a current without spin flip $\tau_{z}$. (There are additional combinations from flips of both spins \cite{horovitz2}, but outside our frequency range and not considered here.)
Hence the current correlations are proportional to
\beq{04}
C_2(\nu_1,\nu_2,\omega) &=& \langle (\tau_-\otimes\tau_z)_t(\tau_+\otimes\tau_z)_0\rangle_\omega
\\
&& {} + \langle (\tau_z\otimes\tau_-)_t(\tau_z\otimes\tau_+)_0\rangle_\omega +(+\leftrightarrow -)
\nonumber
\eeq
where $\langle \ldots \rangle_{\omega}$ is a Fourier transform.
The frequencies $\nu_{1,2} = \nu-a,\nu,\nu+a$ arise from the nuclear quantum numbers $m, m'$ near the first and second spin. Note that the interchange $\nu_1, \nu_2$ yields distinct results, e.g. if $\Gamma_1\neq\Gamma_2$.
For all nuclear configurations of a dimer, the non-degenerate (six pairs) and degenerate (three pairs) contributions are
\beq{11}
C_{\rm nondeg}(\omega)&=&\sum_\pm \big[
C_2(\nu\pm a,\nu,\omega)+C_2(\nu\pm a,\nu\mp a,\omega)
\nonumber\\[-2ex]
&&\qquad {} + C_2(\nu,\nu\pm a,\omega)
\big]
\nonumber\\
C_{\rm deg}(\omega)&=&C_2(\nu-a,\nu-a,\omega)+C_2(\nu,\nu,\omega)\nonumber\\&&\qquad {} + C_2(\nu+a,\nu+a,\omega)
\eeq
The total observable is $C_{\rm nondeg}(\omega) + C_{\rm deg}(\omega)$.

We evaluate the spin-spin correlations via the master equation for both the degenerate and non-degenerate cases and fit parameters to the experimental data. We consider first the general features of the theoretical results, an example is in Fig.\;2(left). The dominant terms are the six non-degenerate terms (higher red lines) that are split by $\pm 2J_{\rm ex}$. Between these six peaks are strong dips at the bare (non-interacting) hyperfine transitions $\nu,\,\nu\pm a$, thus readily determining these parameters.
The degenerate terms (lower black lines) split from the bare transitions by both a strong and a weak splitting $\pm 2(J \pm J_{\rm ex})$ (see Table~I). The weak splitting $2(J - J_{\rm ex}) > 0$ arises from $J_{\rm DM}$, if the latter vanishes, some lines do not split and there would be a (small) peak in the bare locations. Since this was not seen in any of our data we conclude that $J_{\rm DM}$ is significant, i.e.\ at least of order $J_{\rm ex}$. The weak splittings produce side shoulders on the main peaks, seen in some of our data. The strong splitting is larger than that of the non-degenerate terms ($2(J + J_{\rm ex}) > 2 J_{\rm ex}$),
hence, although the overall intensity of the degenerate terms is weaker, the split peaks at the edges of the spectrum become dominant and lead in general to an apparent 8-peak structure.

Before fitting our data, we note that macroscopic ESR for dimers with exchange is totally different from the above. In the macroscopic case a homogeneous rf field is applied so that the measured correlations are symmetric in both spins, $\langle (\tau_-\otimes\tau_z +\tau_z\otimes\tau_-)_t (\tau_+\otimes\tau_z + \tau_z\otimes\tau_+)_0\rangle_\omega$. The permutation symmetry forbids in particular $S\rightarrow T$ transitions for $m=m'$ and $J_{\rm DM}=0$. Furthermore, in the STM case the exchange is generated via the electrodes and is different for degenerate and non-degenerate pairs [compare Eqs.\,(\ref{e02}, \ref{e03})], while in the macroscopic case, there is a direct exchange that applies equally to all $m,\,m'$. Finally, the spin-orbit interaction responsible for the coupling $J_{\rm DM}$ is most likely generated by the heavy metal atoms in the tip, hence absent in a macroscopic setup.

The spectra are limited by the detection sensitivity of the impedance matching circuits to the frequencies $580{-}780\,{\rm MHz}$, thus some of the expected peaks are probably outside this range. We have included in our analysis only data that have at least seven peaks (except for Fig.\;3d, see below); we have found 15 such spectra. Fig.\;2(right) shows the theory spectrum (black smooth line), the sum of the two curves in Fig.\;2(left), compared to experimental data, the magenta jagged curve. The parameters of the fit are given in the caption, in addition we use the linewidths $\Gamma_1, \Gamma_2 \approx 4\,{\rm MHz}$. (We note that in the absence of splitting due to $J_{\rm ex}$ and $J_{\rm DM}$, the linewidth becomes $\Gamma_1 + \Gamma_2$.)

\begin{figure}
\includegraphics*[height=.27\columnwidth]{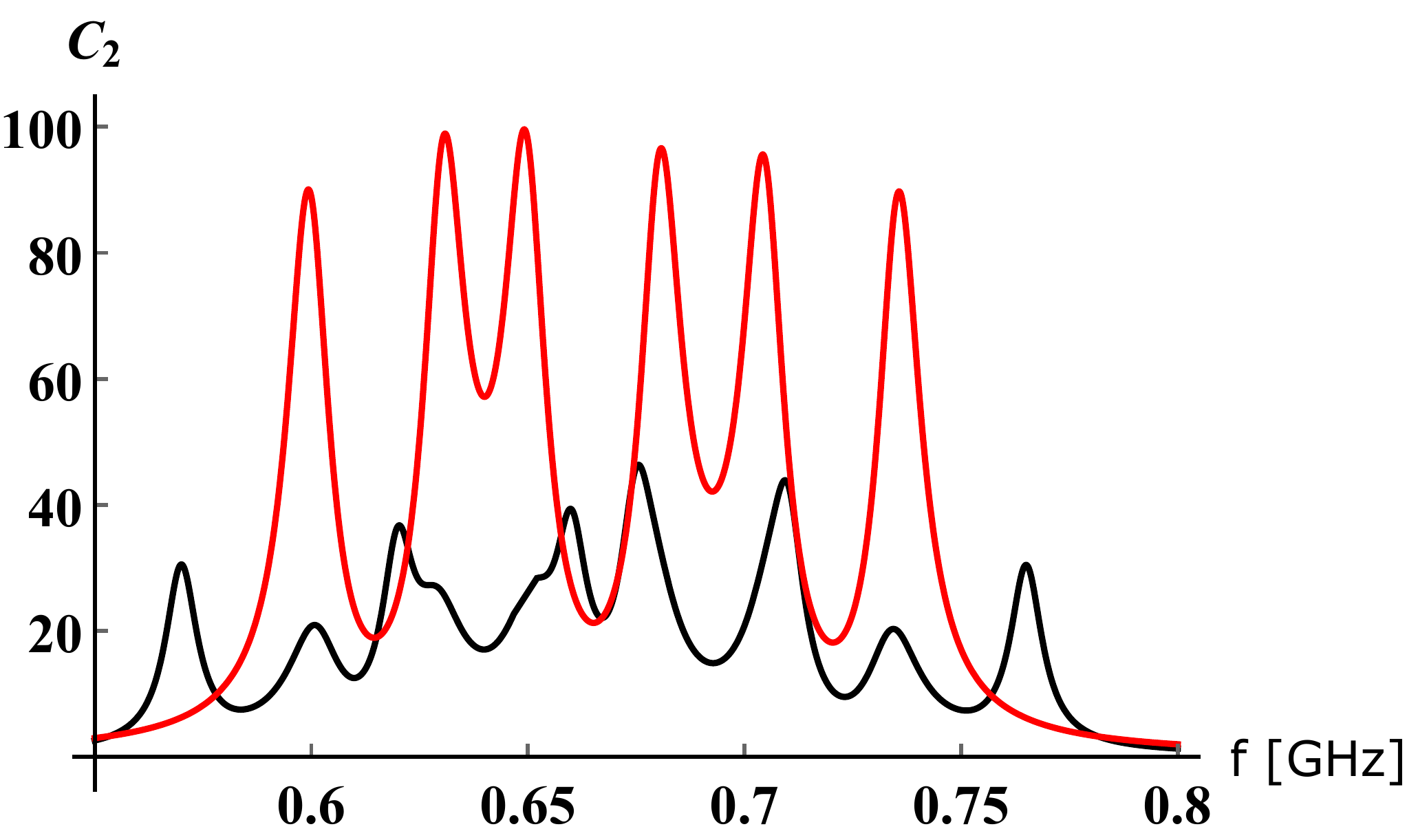}%
\hspace*{0.5mm}
\includegraphics*[height=.27\columnwidth]{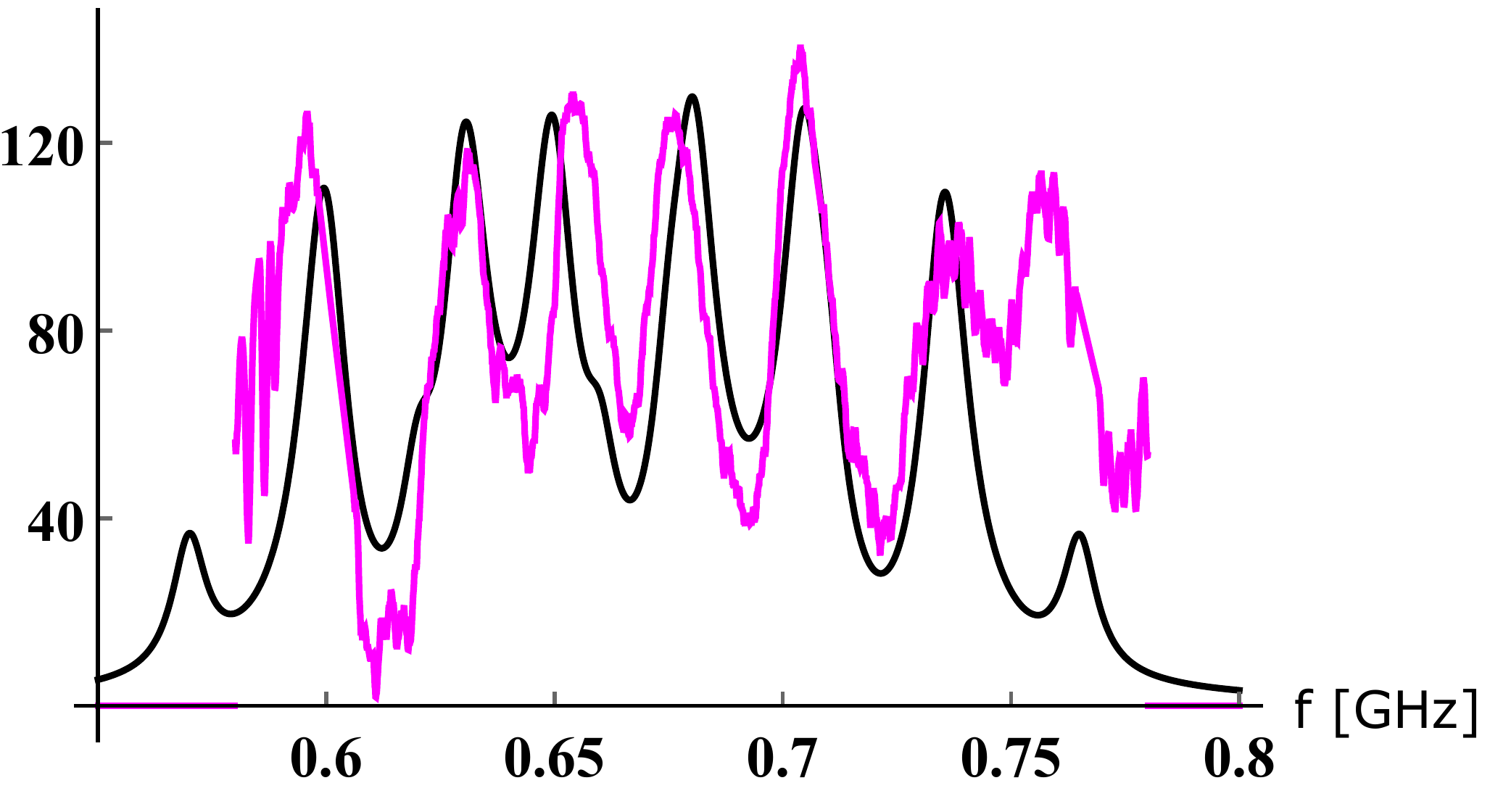}%
\caption{Left: Spectra of nondegenerate correlations (upper red line) and of degenerate ones (lower black line) vs.\ $f = \omega /2\pi$. Right: Experimental (magenta) and theory (black) spectra. Fitting parameters in MHz: $\nu:665$, $a:50,55$, $J_{\rm ex}:8$, $J_{\rm DM}:9$; hyperfine $a$ is somewhat different between left and right.
}
\end{figure}

Fig.\;3(a-c) shows additional fitting curves with fairly similar parameters, obtained for different spots on the sample.
In all cases, reasonable agreement is found with realistic fit parameters, note in particular that the couplings $J_{\rm ex}$, $J_{\rm DM}$ are comparable in magnitude, which points, on the basis of the model, towards a sizable rotation angle $\tan \phi \sim 1$ in the spin-orbit interaction.
The fitted values of $a$ are in between the hyperfine eigenvalues for the $z$ and $x$, $y$ axes \cite{kobayashi}. This suggests some tilt or rotational averaging of the molecular $z$ axis relative to  the magnetic field (normal to the surface).

An exception is Fig.\;3d that appears to have only 3 peaks, as if $J_{\rm ex} = J_{\rm DM} = 0$. Fitting the data this way yields line widths $\Gamma_1,\,\Gamma_2$ that are much larger than all other cases, however.
We believe that it is more likely that a weak, but finite exchange applies in this case (caption of Fig.\;3d), causing shoulders and an apparent increase in width of these lines.

In conclusion, we have found a large set of ESR-STM spectra that that  fits well to a theory of two spins located on a molecular dimer and coupled via electrons that tunnel between tip and substrate. The fitting parameters give effective exchange and Dzyaloshinskii-Moriya couplings that are comparable and fairly strong (comparable to the hyperfine splitting, using the conventional definitions $4J_{\rm ex},\,4J_{\rm DM}$ of Ref.\;\onlinecite{wertz}).
Our analysis of these dimers opens a route for studying hyperfine interactions and g-factors in molecules and determining their parameters. It also paves the road to measure spin-orbit coupling for tunneling electrons.

\bigskip

\begin{figure}[t]
\includegraphics*[height=.27\columnwidth]{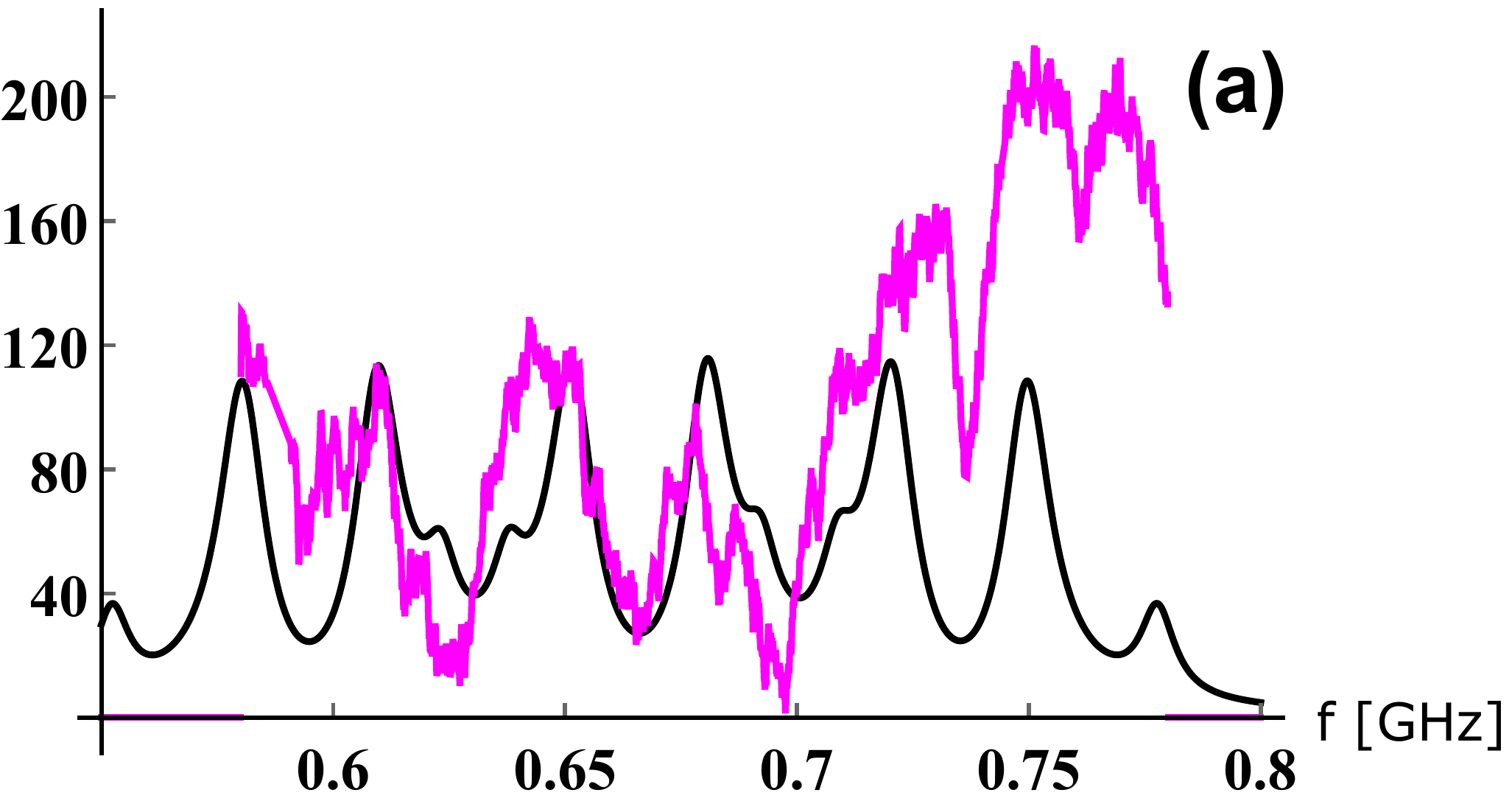}%
\hspace*{0.5mm}
\includegraphics*[height=.27\columnwidth]{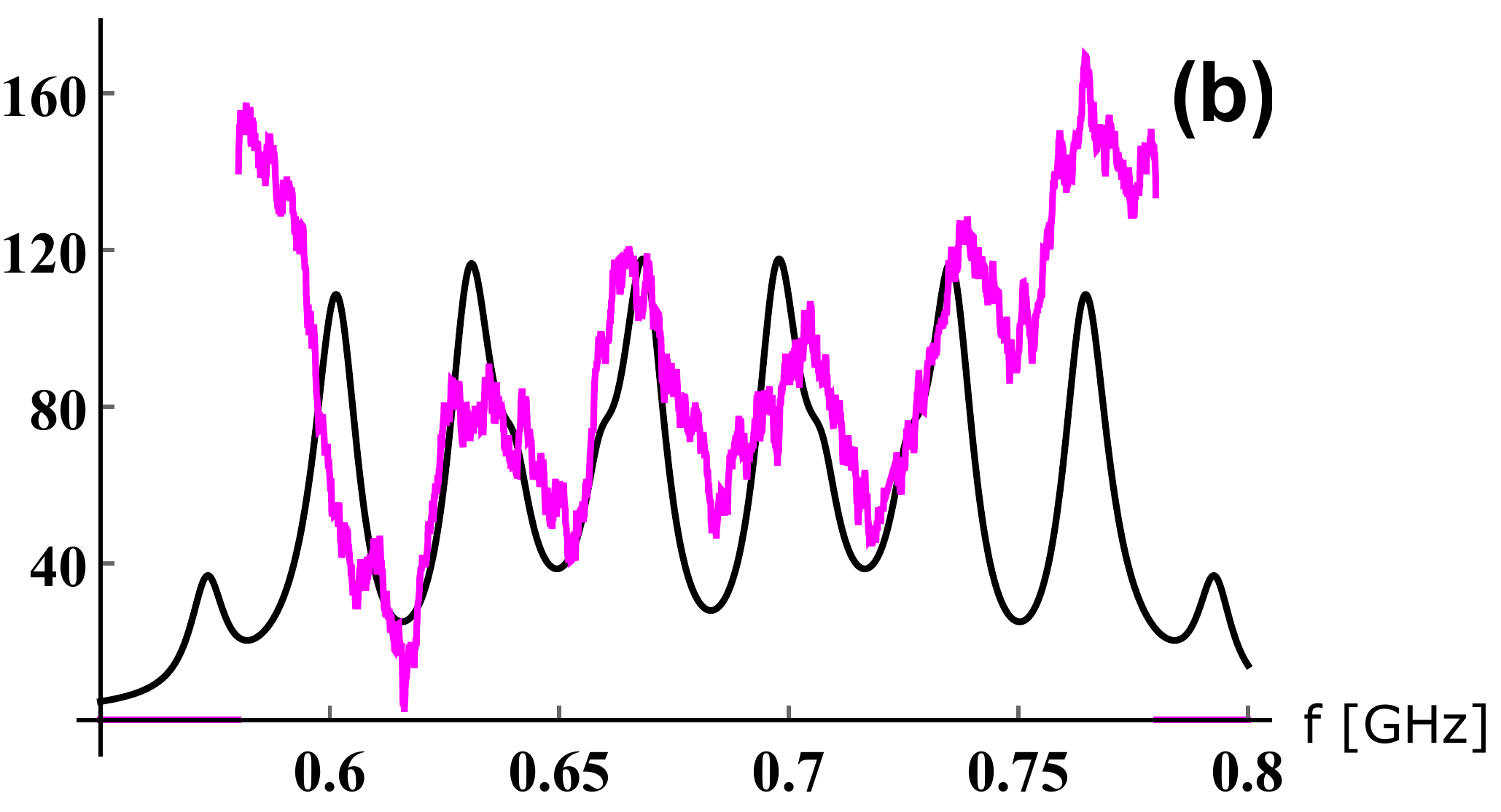}%
\vspace*{1.5mm}
\includegraphics*[height=.27\columnwidth]{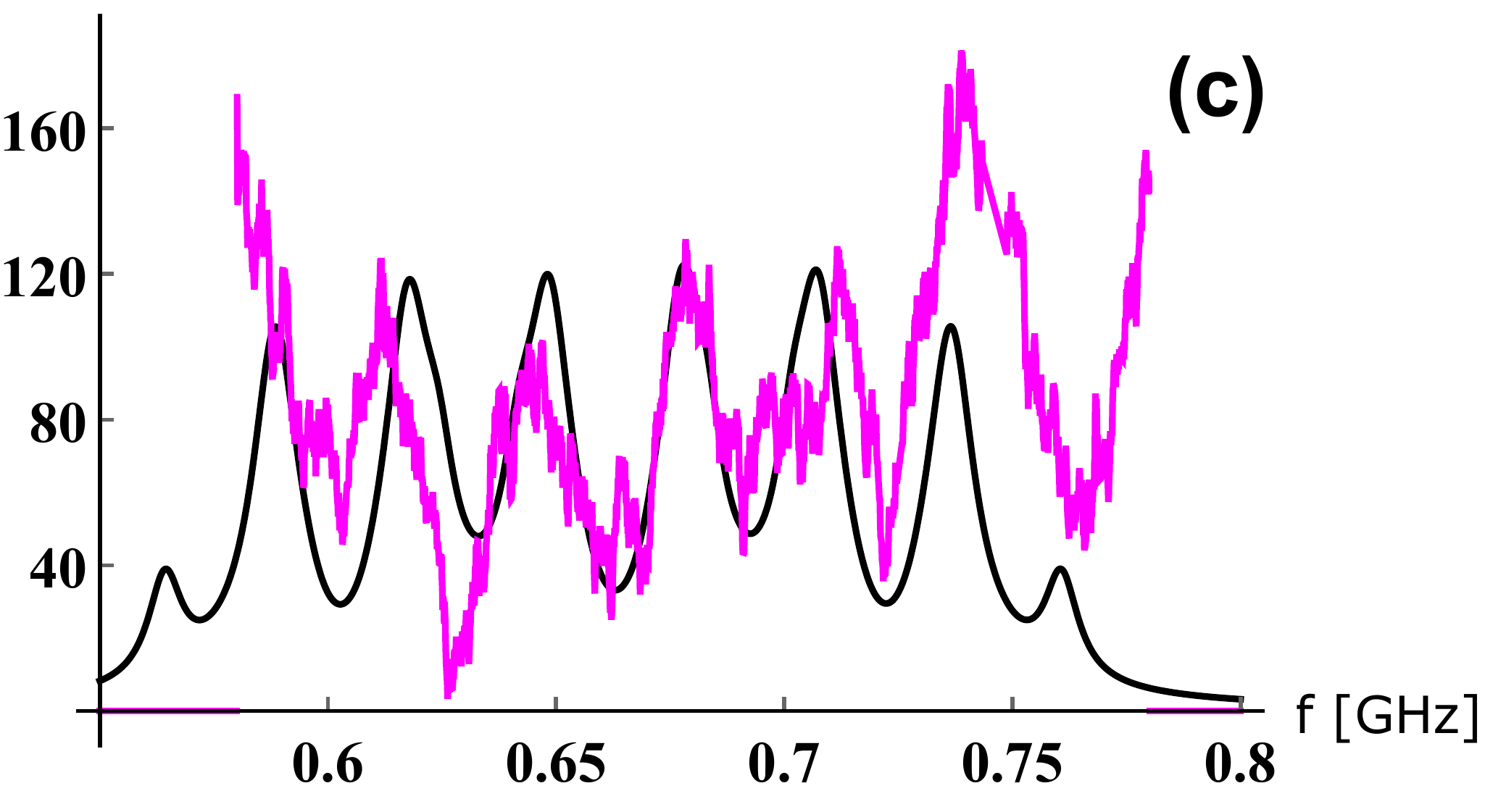}%
\hspace*{0.5mm}
\includegraphics*[height=.27\columnwidth]{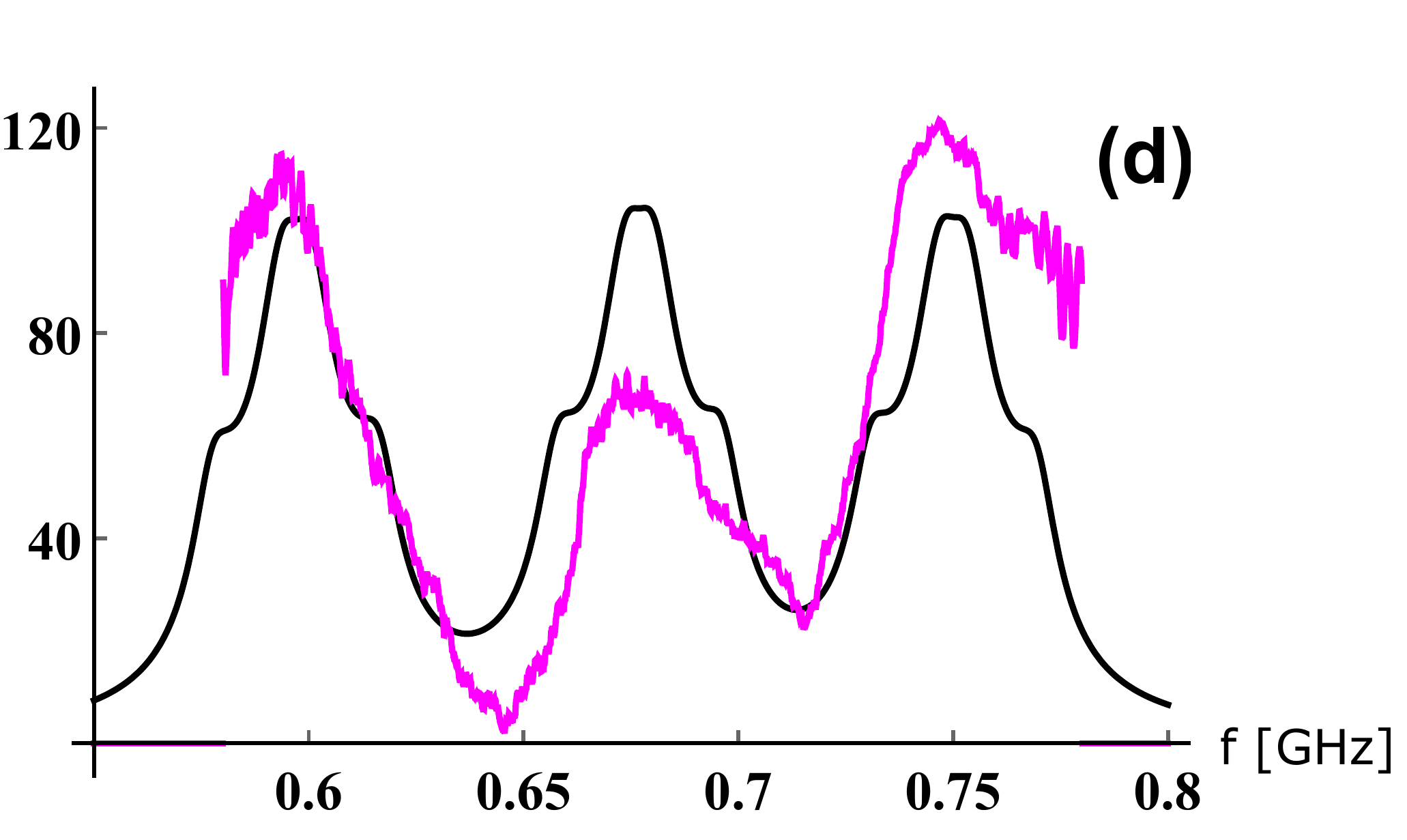}%
\hspace*{0.5mm}
\caption{Experimental ESR-STM spectra (jagged magenta lines) and fitted theoretical curves using the following parameters (in MHz): (a) $\nu:666$, $a:71,\,69$, $J_{\rm ex}:7$, $J_{\rm DM}:12$. (b) $\nu:683$, $a:67$, $J_{\rm ex}:7$, $J_{\rm DM}:12$. (c) $\nu:663$, $a:60,\,59$, $J_{\rm ex}:7$, $J_{\rm DM}:10$. (d) $\nu:677$, $a:80,\,73$, $J_{\rm ex}:1$, $J_{\rm DM}:8$. The small differences in $a$ correspond to different nuclear $m$ levels and may arise from second-order shifts due to sub-leading hyperfine couplings.}
\end{figure}

\acknowledgements This work was funded by Attract -- Developing breakthrough technologies for science and
society, grant NMR(1). Additional funding was provided by the ISF collaboration with China, grant 2811/19, and the German DFG project number 282981637. C.H. and B.H. gratefully acknowledge funding by the German DFG through the DIP programme (Fo 703/2-1).

\end{document}